# Spatiotemporal Dissociation of Brain Activity underlying Subjective Awareness, Objective Performance and Confidence

Abbreviated title: **Slow Cortical Potentials in Conscious Perception**

Qi Li[1], Zachary Hill[1] and Biyu J. He[1*]

[1]National Institute of Neurological Disorders and Stroke, National Institutes of Health, Bethesda, Maryland 20892, USA

[*]Correspondence should be addressed to: Dr. Biyu J He, 10 Center Drive, Building 10, Room B1D728, Bethesda, MD 20892 USA. Tel: +1 (301) 594-0950; Fax: +1 (301) 480-2558; Email: biyu.he@nih.gov

Number of Pages: 49;

Number of Figures: 11;

Number of words for Abstract: 188;

                    Introduction: 498;

                    Discussion: 1,721.

The authors declare no competing financial interests.

**ABSTRACT**

Despite intense recent research, the neural correlates of conscious visual perception remain elusive. The most established paradigm for studying brain mechanisms underlying conscious perception is to keep the physical sensory inputs constant and identify brain activities that correlate with the changing content of conscious awareness. However, such a contrast based on conscious content alone would not only reveal brain activities directly contributing to conscious perception, but also include brain activities that precede or follow it. To address this issue, we devised a paradigm whereby we collected, trial-by-trial, measures of objective performance, subjective awareness and the confidence level of subjective awareness. Using magnetoencephalography (MEG) recordings in healthy human volunteers, we dissociated brain activities underlying these different cognitive phenomena. Our results provide strong evidence that widely distributed slow cortical potentials (SCPs) correlate with subjective awareness, even after the effects of objective performance and confidence were both removed. The SCP correlate of conscious perception manifests strongly in its waveform, phase and power. By contrast, objective performance and confidence were both contributed by relatively transient brain activity. These results shed new light on the brain mechanisms of conscious, unconscious and metacognitive processing.

# INTRODUCTION

The elucidation of brain mechanisms underlying conscious perception requires distilling neural processes directly contributing to conscious awareness from those that precede, follow or co-vary with it. Most investigations on brain mechanisms underlying conscious perception have adopted a "minimal contrast" approach, that is, to hold the sensory inputs as constant as possible and identify brain activities that correlate with the changing content of conscious awareness (Dehaene and Changeux, 2011). However, such a contrast between perceptual conditions would not only reveal the neural correlates of conscious perception (NCC) per se, but also include the prerequisites for (NCC-pr) and the consequences of (NCC-co) conscious perception (Bachmann, 2009; Aru et al., 2012a; de Graaf et al., 2012). An example of the NCC-pr is the gross brain excitability or attentional state at stimulus onset that biases whether the stimulus is consciously perceived or not. Examples of the NCC-co include the rendering of verbal report and construction of long-term memory.

We aimed to address this issue by controlling for behavioral variables correlated with conscious perception. Combining liminal stimulation, forced alternative-choice and trial-by-trial introspection (Sergent et al., 2005; Wyart and Tallon-Baudry, 2008; Lamy et al., 2009; Fleming et al., 2010; Rounis and Lau, 2010; Hesselmann et al., 2011), we devised a paradigm in which we collected, on every single trial, measures of objective performance, subjective awareness, and confidence level about subjective awareness (Fig. 1A). The mutual correlations between these behavioral measures suggest that a simple experimental contrast of subjective awareness (seen vs. unseen) would likely include brain activities underlying objective performance and confidence as well. However, the trial-by-trial collection of all three behavioral measures allowed us to disentangle brain activities associated with each of them. Our reasoning is as follows: since the NCC-pr affects both subjective awareness and objective performance (Lamme, 2003; Dehaene et al., 2006) while the NCC only contributes to subjective awareness (by definition), including the factor of objective performance in our design should help separating the NCC from NCC-pr. Similarly, since metacognitive processes are a component of the NCC-co (Aru et al., 2012a), including confidence measure in our paradigm should facilitate isolating the NCC from NCC-co.

We combined this behavioral paradigm with MEG recording to investigate brain activities underlying conscious perception, with the correlated effects of objective performance and confidence removed. We further aimed to test the SCP hypothesis of conscious awareness (He and Raichle, 2009b). This idea postulates that the SCPs – the low-frequency (<4 Hz) component of brain field potentials (He et al., 2008) – might be a correlate of conscious awareness. Previous studies have established that long-lasting excitatory postsynaptic potentials (EPSPs) at the apical dendrites of pyramidal neurons in the superficial layers are the major contributor to surface-recorded SCPs (Mitzdorf, 1985; Birbaumer et al., 1990; He and Raichle, 2009b). While the SCP is traditionally recorded with scalp-electroencephalography (EEG) (Birbaumer et al., 1990) or surface-electrocorticography (ECoG) (He et al., 2008), it has recently been shown that the SCP can also be recorded using MEG (Brookes et al., 2005; Leistner et al., 2007).

**MATERIALS AND METHODS**

**Subjects**

The experiment was approved by the Institutional Review Board of the National Institute of Neurological Disorders and Stroke. All subjects were right-handed, neurologically healthy with normal or corrected-to-normal vision. Eleven subjects between 22 and 38 years of age (mean age 27; six females) participated in the first phase of the study consisting of a single MEG session lasting ~3 hours. Two additional subjects (both female, age 24 and 25) performed three MEG sessions and one EEG session on separate days using an identical task (see **Assessment of Test-retest Reliability** and **EEG Data Collection**). Lastly, a second cohort of 11 subjects (age range: 23 to 39, mean 25 years; seven females) each participated in a session of behavioral testing with a modified version of the task (see **Behavioral Control for the Length of Post-stimulus Blank Period**). All subjects provided written informed consent.

**Stimuli and Task**

The stimulus was shown on a Panasonic DLP Projector (PT-D3500U, refresh rate: 60 Hz) with a 10% neutral density optical filter in front of the lens. The filter was applied to make the luminance as low as 1 $cd/m^2$ on the stimulus screen, such that every subject could reach a threshold duration longer than 16.7 ms – the limitation of the projector refresh rate. The stimulus was presented on a screen 75 cm away from the subject's eyes. Dark adaptation (at least 30 minutes) was conducted for all 11 subjects before any behavior data were collected.

Each trial started with a white fixation cross on a gray background (Fig. 1A). When the subject was ready, s/he pressed a button to start the trial. After a blank screen of a random duration between 2 and 6 sec (following an exponential distribution), a Gabor patch (1° visual angle/cycle) with a very low contrast (1%) was presented for a short duration (see below). The orientation of the Gabor patch was randomly selected between 45° and 135° with equal chance. Then another blank screen with a duration randomly chosen between 3 and 6 sec (following an exponential distribution) was presented. The luminance of the blank screens was equal to the background luminance of the stimulus screen. The first blank period ensured that the subject could not predict the onset of the stimulus. The second blank period allowed enough time after stimulus offset for the analyses of slow MEG activity uncontaminated by responses to questions. Each trial ended with three sequential questions: i) A forced alternative-choice – Was the Gabor patch pointing to upper-left (135°) or upper-right (45°)? ii) Did you see the stimulus (Gabor patch) or not? iii) On a scale of 1 to 4, how confident are you about your answer to question ii, with level 4 indicating "absolutely sure" and level 1 "not sure at all". The three questions were designed to probe objective performance (OBJ), subjective awareness (SUB) and confidence level (CONF), respectively. Subjects indicated their answers to the questions via a fibreoptic key-pad (LumiTouch).

The experiment was conducted in two stages. In the first stage, the duration of the Gabor patch was adjusted using Levitt's staircase method (Levitt, 1971), until an individually titrated threshold for subjective awareness was found (i.e., the subject answers "seen" to Question ii in about half of the trials). The full trial as described above was used in this stage to familiarize the subject with the different questions. The distribution of threshold duration across subjects was as follows: six subjects: 33.3 ms; three subjects: 50 ms; two subjects: 66.7 ms. Once the threshold duration was determined, in the second stage of the experiment, trials were shown repeatedly with identical stimulus duration at the subject's individual threshold while MEG signals were continuously recorded. Subjects performed these trials in sessions of <12 min long and were allowed to rest between sessions. In total, 113 ~ 277 trials were acquired in each subject (mean±s.d. across subjects: 183.5±56.4) after artifact rejection. Additionally, 6 ~ 10 catch trials without a stimulus were shown for each subject. All of the reported behavioral and MEG results are from the second stage of the experiment only.

**Data Acquisition**

Experiments were conducted in a whole head 275-channel CTF MEG scanner (VSM MedTech, Coquitlam, Canada). MEG data were collected under a sampling rate of 600 Hz with an anti-aliasing filter at <150 Hz. Before and after each recording session, the head position of the subject was measured with respect to the MEG sensor array, using coils placed on the ear canals and the bridge of the nose. All MEG data samples were corrected with respect to the refresh delay of the projector (measured with a photodiode). In 10 out of 11 subjects, anatomical magnetic resonance imaging (MRI) data were acquired on a General Electric 3T scanner with an 8-channel head coil, using a MP-RAGE sequence with a resolution of 1×1×1 mm$^3$. The MEG data were aligned to the anatomical MRI, using the coils placed at the anatomical landmarks.

**Event-related Fields (ERFs) and Sensor-space Analysis of Variance (ANOVA)**

The Fieldtrip package (http://fieldtrip.fcdonders.nl) implemented in Matlab (Mathworks Inc.) was used for data analysis. First, the whole recording from each session was detrended, mean-removed, and bandpass filtered at 0.05 ~ 35 Hz with a 4$^{th}$ order Butterworth filter (all filtering was done offline with an acausal filter). Second, data were epoched from 1 sec before the stimulus onset to 3 sec after. Third, independent component analysis (ICA) was performed to remove eye movement and cardiac artifacts (Bell and Sejnowski, 1995), and trials with artifacts were rejected manually. Fourth, baseline-correction was carried out using a window of -500 ~ 0 ms.

The resulted dataset were subjected to ERF analysis, sensor-space and source-space ANOVA. ERF analysis consisted of averaging across trials defined as the same type (e.g., seen vs. unseen). For sensor-space ANOVA, the data were down-sampled to 100 Hz before a trial-by-trial three-way ANOVA (factors: SUB, OBJ and CONF; dependent measure: MEG activity) was conducted at every sensor location and time point.

**Source Localization**

Source localization was conducted using the cortically constrained L2 minimum-norm estimate (MNE) implemented in Fieldtrip for the ten subjects in whom anatomical MRI data were collected. First, brain surface for each subject was segmented and registered to the spherical atlas in Freesurfer (http://surfer.nmr.mgh.harvard.edu/fswiki). Then the mesh points of the surface were decimated using the MNE suite (http://www.martinos.org/mne/) to obtain the same number of nodes on the cortical surface for each individual (N = 4098 for each hemisphere). A single-shell model was applied to each individual's cortical surface to construct the forward model.

Standard source localization methods are typically applied to trial-averaged ERFs [(Dale and Sereno, 1993; Hämäläinen, 2005) but see (Brookes et al., 2011)]. However, in our case, the correlation between different behavioral measures precludes the ERF as an accurate reflection of brain activity unequivocally underlying each behavioral measure. To resolve this issue, we constructed the source estimate for each trial individually and performed a trial-by-trial analysis in the source space. In other words, the source estimate of each trial was obtained without information from other trials, so as not to introduce correlation among behavioral factors in the source space and to allow unbiased trial-by-trial analysis. For the estimation of noise covariance matrix at the single-trial level, we used the entire time course of the full epoch (-1 ~ 3 sec) of each trial, instead of just the pre-stimulus window, to achieve a more stable estimate (Brookes et al., 2011). We performed a control analysis to compare the noise covariance matrices estimated from the pre-stimulus window and from the full epoch, and found them to be highly similar (all $P < 1e-16$, assessed by spatial correlation of the half-matrix). Lastly, current estimates were transformed into an approximately normal distribution (using *boxcox* function in matlab) (Yao and Dewald, 2005; Watthanacheewakul, 2010), and a three-way ANOVA (factors: SUB, OBJ and CONF; dependent measure: source activity estimate) was carried out at every source location in each subject to isolate the effect due to each behavioral measure. Source-space ANOVA was performed at selected time points from 100 ms to 2.5 sec after stimulus onset.

For population analysis, results from each subject were smoothed using a 6-mm-width kernel. The percentage of subjects showing a significant effect at each source location and time point was then calculated and subjected to binomial statistics (He, 2013; He and Zempel, 2013).

**Frequency-domain Analyses**

For frequency-domain analyses, MEG data were epoched from 2 sec before the stimulus onset to 3 sec after. A longer pre-stimulus window was used to allow better estimation of low-frequency phase and power. For the analysis on phase, MEG data were filtered in 20 frequency bands: [0.05, 1], [1, 3], [3, 5], [5, 7] … [37 39] Hz using a 3$^{rd}$ order Butterworth filter. Then, in each trial, instantaneous phase was extracted from each frequency band using Hilbert transform. The phase time series from each sensor was down-sampled to 50 Hz before a trial-by-trial two-way circular ANOVA was conducted at each time-frequency location in each subject (factors: SUB and OBJ; dependent variable: phase). A two-way instead of three-way circular ANOVA was used because statistical methods for circular data are still in development (Berens, 2009). All circular statistics were carried out using the CircStat toolbox implemented in Matlab (Berens, 2009).

For the analysis on power, a Gabor wavelet $G(t,f) = \exp\left(-\frac{t^2}{2\sigma_t^2}\right) \exp(j2\pi f t)$ with $\sigma_t = \frac{5}{f}$ was used. The following center frequencies were analyzed: from 2 to 30 Hz with 1-Hz steps; from 31 to 61 Hz with 2-Hz steps; and from 63 to 148 Hz with 5-Hz steps. 2 Hz was the lowest frequency analyzed because accurate estimation of power requires sufficient length of data, which was limited by the length of the trial. Power time series from each trial were log-transformed into roughly normally distributed data (Wyart and Tallon-Baudry, 2008), and baseline corrected using a window of -750 ~ -550 ms. A trial-by-trial three-way ANOVA was conducted at every time-frequency-sensor location in each subject (factors: SUB, OBJ and CONF; dependent variable: MEG power). In addition, a two-sample t-test was carried out for the effect of subjective awareness (seen vs. unseen) and objective performance (correct vs. incorrect) respectively.

For group analysis, significant sensors from the ANOVA (for the effect of SUB or OBJ, at a P < 0.05 level) were pooled together by summing the t-score from the corresponding t-test across sensors [$\Sigma_T$, see (Wyart and Tallon-Baudry, 2008) for the use of a similar metric]. To assess whether positive and negative t-scores might be cancelled out during this process, we performed a control analysis: At each time-frequency location, sensors with positive and negative t-scores were summed together separately. The results from this analysis confirmed that positive and negative t-scores were largely separate in the time-frequency space, suggesting that cancellation effect was minimal.

The $\Sigma_T$ metric indexes the total level of "seen vs. unseen" or "correct vs. incorrect" contrast across the brain. Because only significant sensors from the ANOVA were included, correlations between different behavioral measures were minimized. $\Sigma_T$ is a normally distributed variable and was subjected to a one-sample t-test across the 11 subjects ($H_0$: $\Sigma_T = 0$). Lastly, the population-level results were assessed for statistical significance using a nonparametric permutation test, which effectively controls the type I error (i.e., the false alarm rate) in a situation of multiple comparisons by clustering neighboring time-frequency points that exhibit the same effect (Nichols and Holmes, 2002; Jokisch and Jensen, 2007; Medendorp et al., 2007).

**Distribution of Preferred Phase in Seen vs. Unseen Trials**

This analysis was applied to the phase of the lowest frequency band (0.05 ~ 1 Hz) at stimulus onset only.

*Cluster-level analysis*

For each subject, P-values from the two-way circular ANOVA on phase (factors: SUB, OBJ) for the effect of subjective awareness were plotted on the scalp. In all subjects, significant sensors formed spatial clusters. We adjusted the P-value threshold (all $P < 0.001$) to define at least two well-separated clusters in each subject. The two largest clusters in each subject were extracted. For each cluster thus defined, we pooled across all sensors within the cluster and all seen (or unseen) trials. A circular histogram of phase was constructed for seen and unseen trials separately, using 10 evenly distributed phase bins. The histogram shows the fraction of trials falling into each bin. The circular-mean of this histogram was computed to obtain the averaged vector. The norm of this vector defined the phase-locking value (PLV) (Tallon-Baudry et al., 1996; Lachaux et al., 1999) and the angle of this vector defined the preferred phase. The PLV provides a direct index of the concentration of phase across trials, and the preferred phase describes the most common phase among the group of trials analyzed.

*Sensor-level analysis*

All sensors with a significant effect for subjective awareness (at a $P < 0.001$ level) from the two-way ANOVA on phase were included in this analysis. For each sensor, the phase from each trial was described as a unit vector with the corresponding angle. This vector was circular-averaged across trials of the same type (seen vs. unseen) to obtain the averaged vector, the amplitude and angle of which then defined the PLV and the preferred phase of this particular sensor.

**Analysis on Confidence**

Because ANOVA cannot reveal brain activity monotonically related to confidence, we used an alternative method – ordinal logistic regression (De Martino et al., 2013). Confidence ranking (values: 1, 2, 3, 4) was an ordinal variable treated as the dependent measure. Independent measures included the estimated source activity (continuous variable), subjective awareness and objective performance (binary variables). The regression model is as follows:

$$\log\frac{P(c \leq i)}{1-P(c \leq i)} = a_i - \beta_{act}S_{act} - \beta_{sub}A_{sub} - \beta_{obj}A_{obj},$$

where $c$ is the confidence level and $i$ takes values of 1, 2 and 3. $S_{act}$ is the source activity, $A_{sub}$ and $A_{obj}$ are the answers to the subjective awareness (0 – unseen; 1 – seen) and objective performance (0 – incorrect; 1 – correct) questions respectively, and $\beta_{act}$, $\beta_{sub}$ and $\beta_{obj}$ are the associated regression coefficients. The left hand of the above equation is called a logit function. For each logit function (associated with a particular $i$ value) there was a different intercept term $a_i$, but the regression coefficients $\beta_{act}$, $\beta_{sub}$ and $\beta_{obj}$ were the same across the three logit functions. Positive $\beta$ values indicate a positive correlation, i.e., higher confidence levels were associated with a larger predictor value. The *mnrfit* function in Matlab was used to carry out the ordinal logistic regression.

For group analysis, we first computed the number of subjects showing a significant (P < 0.05) $\beta_{act}$ at each source location. Only source locations with at least two subjects showing a significant result (corresponding to population P < 0.07, binomial statistics) were analyzed further. We then computed an agreement coefficient showing the degree of agreement across subjects in terms of the sign of $\beta_{act}$. To this end, we first determined the number of subjects with a significant positive or negative $\beta_{act}$, designated *Npos* and *Nneg* respectively. The agreement coefficient (AC) was calculated as follows:

$$\text{If } Npos > Nneg, \; AC = Npos / (Npos + Nneg);$$
$$\text{If } Npos < Nneg, \; AC = -Nneg / (Npos + Nneg);$$
$$\text{If } Npos = Nneg, \; AC = 0.$$

In order to accurately estimate the agreement coefficient, single-subject results were not subjected to spatial smoothing in this analysis.

**Assessment of Test-retest Reliability and Control for Signal-to-Noise Ratio (SNR)**

In order to assess test-retest reliability, we collected a substantially larger amount of data in two additional subjects (Subj. #12 and #13). Each of these subjects completed three experimental sessions on three different days, with each session lasting ~3 hours. The task design was the same as described above. In total, 1010 and 890 trials were recorded in the two subjects (after artifact rejection), respectively. For each session, the MEG data were analyzed for sensor-space ERF and ANOVA according to procedures described above. Furthermore, in order to control for the unbalanced number of trials across behavioral conditions, which results in differential SNR, we equated the number of trials across behavioral conditions in an additional ERF analysis by randomly dropping out a fraction of trials (these trials were evenly distributed throughout the session). Because the head position relative to MEG sensors varied across experimental sessions, each session was analyzed separately.

**EEG Data Collection and Analysis**

The same two subjects (Subj. #12 and 13) each participated in an additional EEG session under an identical task design. EEG data were recorded using a 64-channel DC-EEG system (BrainAmpDC, Brain Products GmbH) with a mastoid reference in an EMI-shielded, sound and lighting controlled room. Vertical and horizontal electrooculogram (EOG) were simultaneously recorded. Analysis steps were as follows: 1) rejecting channels and data segments contaminated by artifacts via inspection of the raw data records; 2) filtering between 0.05 and 150 Hz; 3) re-referencing to a linked-mastoid reference; 4) removal of eye blink and eye movement artifacts by an ocular correction ICA (Infomax Extended ICA implemented in Analyzer2, Brain Products GmbH); 5) epoching from 1 sec before to 3 sec after stimulus onset; 6) baseline correction using a window of -500 ~ 0 ms; and 7) averaging across trials to yield event-related potentials (ERPs). 300 trials were recorded in each subject, including 10% catch trials.

**Behavioral Control for the Length of Post-stimulus Blank Period**

So far, our paradigm has used a 3~6 sec post-stimulus blank period in order to extract slow brain activity uncontaminated by the behavioral responses. In order to control for the potential decay of working memory during this period, we conducted behavioral testing in an additional cohort of 11 subjects using a 200-ms-long post-stimulus blank period under an otherwise identical behavioral paradigm. Testing was conducted in a sound and lighting controlled room outside the MEG scanner. 314±48 (mean±SEM) trials were recorded in each subject, including 10% catch trials.

## RESULTS

**Behavioral Results**

The task design is illustrated in Fig. 1A and explained in detail in Materials and Methods (section **Stimuli and Task**). Subjects reported seeing the stimulus in 48.9±4.4% (mean±SEM across subjects) of trials, suggesting that the threshold for subjective awareness was successfully reached. In addition, they answered correctly about the orientation of the Gabor patch in 79.5±2.5% of trials (Fig. 1B, chance level = 50%). When subjects reported seeing the stimulus, they correctly identified the orientation of the Gabor patch in 96.8±1.3% trials (Fig. 1C). Interestingly, even when they denied seeing the stimulus, their objective performance remained above chance level (%correct: 62.0±2.7%, $P = 0.001$, assessed by a Wilcoxon signed rank test against 50%). This is consistent with the "blindsight" phenomenon reported in many previous studies, i.e., even when the subject denies seeing the stimulus, their performance in a forced alternative-choice question can be well above chance level (Weiskrantz, 2004; Lau and Passingham, 2006; Del Cul et al., 2009; Hesselmann et al., 2011). Since subjects' performance in orientation discrimination was much better in "seen" than "unseen" trials (Fig. 1C), objective performance and subjective awareness were indeed correlated from trial to trial.

When we considered objective performance against both subjective awareness and confidence (Fig. 1D), we found that confidence has opposite relationships with objective performance in seen vs. unseen trials. In seen trials, the more confident the subject was about having seen the stimulus, the better the objective performance. In unseen trials, the more confident the subject was about *not* seeing the stimulus, the lower their objective performance. A two-way ANOVA indicated a significant effect of both factors and their interaction (SUB: $P < 1e-15$; CONF: $P = 0.002$; SUB × CONF: $P = 1e-5$). Considering subjective awareness against objective performance and confidence (Fig. 1E), we found a significant effect of objective performance ($P < 1e-8$, assessed by a two-way ANOVA), a trend effect for confidence ($P = 0.08$), and a non-significant interaction effect (OBJ × CONF: $P > 0.6$).

**Event-Related Fields**

We first investigated how trial-averaged ERFs vary with objective performance and subjective awareness. Separating all trials according to subjective awareness, we found that compared to unseen trials, seen trials were characterized by dramatically enhanced long-lasting ERFs starting from ~300 ms after stimulus onset and persisting for 2~3 sec, spanning the entire duration of the trial. Data from an example subject (Subj. #1) is presented in Fig. 2A (top panels); similar qualitative results were obtained in all subjects. Importantly, this long-lasting activity is not of an oscillatory nature but rather is a slow DC-type drift – a signature of the SCP. In Subj. #1, there was also an ERF peak at ~250 ms that was more pronounced in seen than unseen trials; however, this peak was less reproducible across subjects. Separating all trials according to correct vs. incorrect objective performance yielded similar results, although the magnitude of the long-lasting activity in correct trials was smaller than that in seen trials (Fig. 2A, bottom panels). However, because subjective awareness and objective performance were correlated from trial to trial (Fig. 1C-E), the similarity between ERFs of seen trials and correct trials could simply be due to a fraction of shared trials between them.

Indeed, when we sorted all trials according to the combination of subjective awareness and objective performance, we found that only trials that were both seen and correct still contained long-lasting ERFs (Fig. 2B, data from Subj. #1). Unseen trials, whether correct or incorrect, did not contain long-lasting ERFs (Fig. 2B). For group analysis, we performed grand average on the absolute value of the MEG activity time course across all sensors and all subjects, for the three task conditions separately: Seen and correct; Unseen and correct; Unseen and incorrect (Fig. 2C). Because on average 96.8% of seen trials had correct objective performance, there were not enough trials in the "Seen and incorrect" condition for a separate ERF analysis. As can be seen in Fig. 2C, under identical objective performance ("correct"), long-lasting MEG activity is only present in "seen" trials but not "unseen" trials. By contrast, under identical subjective awareness condition ("unseen"), there was no difference in the long-lasting activity between correct and incorrect trials. These results suggest that long-lasting MEG activity is associated with the state of subjective awareness but not objective performance. While the above results are qualitative, to quantitatively dissociate brain activities underlying subjective awareness and objective performance, and to control for the effect of confidence, we next performed a trial-by-trial factorial analysis.

**Trial-by-trial ANOVA on MEG Activity**

A three-way ANOVA, with trial-to-trial MEG activity as the dependent variable and the three independent factors being OBJ, SUB and CONF, was carried out on the MEG activity from every sensor in each subject. The sensor-time locations at which MEG activity correlated with subjective awareness or objective performance in three representative subjects are shown in Fig. 3A. To present the full data set, we used a liberal threshold here ($P < 0.05$, uncorrected). Long-lasting MEG activity still correlated with subjective awareness after the effects of OBJ and CONF were both controlled for (Fig. 3A left). By contrast, after the effects of SUB and CONF were removed, long-lasting activity no longer correlated with objective performance (Fig. 3A right). Similar qualitative patterns were observed at the single-subject level in the remaining subjects. These results are in line with the ERF results reported above. Because the sign of the MEG signal depends on the geometry of the underlying dipole current flow in relation to sensor position and orientation, whether seen trials had higher activity than unseen trials in a particular sensor (warm colors in Fig. 3A) or vice versa (cool colors) was relatively arbitrary.

For population analysis, we compared the number of sensors showing a significant correlation to subjective awareness ($P < 0.05$ for SUB effect, from the three-way ANOVA) with those showing a significant correlation to objective performance ($P < 0.05$ for OBJ effect) at each time point across subjects (Fig. 3B-C). The number of sensors correlated with objective performance stayed relatively flat throughout the trial. By contrast, the number of sensors correlated with subjective awareness abruptly increased at ~200 ms; thereafter it reached peak at ~500 ms and slowly returned to baseline over 2 ~ 3 sec. This result was robust to the particular threshold used for determining significant sensors ($P < 0.01$, Fig. 3D; $P < 0.001$, Fig. 3E).

**Source Localization**

To localize brain activities underlying subjective awareness and objective performance respectively, we performed source localization on single-trial MEG data in the ten subjects with anatomical MRI data. A three-way ANOVA (factors: OBJ, SUB and CONF; dependent variable: estimated source activity) was carried out at every source location in each subject. We focus on the effects of SUB and OBJ here, which show the estimated source activity significantly correlated with subjective awareness or objective performance respectively, each with the other two behavioral factors controlled for. Fig. 4 plots the percentage of subjects at each source location and time point that showed a significant ($P < 0.05$) result for SUB or OBJ effect, thresholded at 40% ($P < 0.001$, uncorrected). The highest level of overlap across subjects was 90% for subjective awareness (corresponding to $P < 2e-11$) and 80% for objective performance (corresponding to $P < 1.6e-9$). Due to the assumptions made in solving the inverse problem of MEG source modeling, statistics in the source space should be considered as "descriptive" instead of veridical.

As can be seen in Fig. 4, brain activity correlated with subjective awareness or objective performance appeared in the occipital cortex at ~200 ms. It spread anteriorly into frontoparietal and temporal cortices by ~300 ms. The most pronounced difference between subjective awareness and objective performance resides between 500 ms and 1.5 sec, when brain activity correlated with subjective awareness covered widespread frontoparietal and temporal areas, while few source locations correlated with objective performance. By about 2 ~ 2.5 sec, widespread brain activity again correlated with both subjective awareness and objective performance, likely in anticipation of the upcoming responses to questions.

**Phase of Slow MEG Activity Correlates with Subjective Awareness**

Thus far we have shown that long-lasting MEG activity in the time domain correlates with subjective awareness. We next investigated the frequency-domain characteristics of this activity. Inspired by earlier studies (Busch et al., 2009; Mathewson et al., 2009), we first examined whether the phase of the MEG activity correlated with subjective awareness or objective performance, each with the other behavioral factor controlled for. For MEG activity from each sensor in each trial, Hilbert transform was used to extract the instantaneous phase in 20 frequency bands covering the range from 0.05 to 39 Hz. We then performed a trial-by-trial two-way circular ANOVA (factors: SUB and OBJ) on the phase of MEG activity at every time-frequency location. The results of this analysis from three representative subjects are shown in Fig. 5A. At each time-frequency location, the number of sensors showing a significant ($P < 0.001$, uncorrected) effect of SUB (left) or OBJ (right) is plotted as color. Significant correlations between MEG activity phase and subjective awareness occurred in the lowest frequencies (0.05 ~ 5 Hz), with the strongest effect in the 0.05 ~ 1 Hz band (Fig. 5A, left).  Much weaker correlations were found between MEG activity phase and objective performance (Fig. 5A, right).  Similar qualitative results were obtained in all subjects.

For group analysis, we conducted a paired t-test across subjects on the number of sensors showing a significant correlation between MEG activity phase and subjective awareness against those showing a correlation between phase and objective performance ($P < 0.001$ for the effect of SUB or OBJ respectively, from ANOVA).  Only the lowest frequencies (<5 Hz) had a significant difference (Fig. 5B).  In particular, the 0.05 ~ 1 Hz band was the only frequency range showing a significant difference at and before the stimulus onset.  Changing the threshold for determining significant sensors from $P < 0.001$ to $P < 0.01$ or $P < 0.05$ yielded very similar results (not shown).

**The Distribution of SCP Phase in Seen vs. Unseen Trials**

The above results show that the phase of the SCP correlates with subjective awareness even after the effect of objective performance was removed.  This raises the following question: At which phase(s) of the SCP is stimulus presentation more likely to be consciously perceived?  To answer this question, we focused on the phase of 0.05 ~ 1 Hz activity at stimulus onset.  Fig. 6A shows MEG sensor locations with a significant correlation between phase and subjective awareness (assessed by the ANOVA) in an example subject (Subj. #1).  It can be seen that significant sensors formed spatial clusters. For the two largest clusters, the distribution of phase across seen and unseen trials are shown in Fig. 6B. While the phase in unseen trials was distributed uniformly around the circle (blue lines), in seen trials it concentrated around $\frac{\pi}{4}$ and $-\frac{3\pi}{4}$ in the two clusters, respectively. Phase $\frac{\pi}{4}$ and $-\frac{3\pi}{4}$ are opposite phases to each other, suggesting that the underlying dipole current flow is likely located between these two clusters of sensors in a perpendicular direction.

We next extracted two dominant clusters of sensors based on the correlation between phase (0.05 ~ 1Hz at stimulus onset) and subjective awareness for each subject in a similar manner (number of sensors included in each cluster: 12.8±2.8), and calculated their phase histograms for seen and unseen trials separately. From the phase histogram we obtained two metrics: i) a phase-locking value (PLV) describing the degree of phase concentration across trials, with value 0 indicating a uniform distribution and value 1 indicating identical phase value across trials; and ii) a preferred phase describing the most commonly observed phase. For the 22 clusters, PLV is plotted against the preferred phase for seen (red) and unseen (blue) trials respectively (Fig. 6C). The degree of phase concentration was significantly higher in seen trials than unseen trials (paired t-test on PLV across clusters, $P < 1.6e-12$). Moreover, in unseen trials, the distribution of preferred phase across the 22 clusters did not depart from uniformity (Rao's test for circular uniformity, $P > 0.5$), whereas the preferred phase in seen trials followed a bimodal distribution, concentrating around $\frac{\pi}{4}$ and $-\frac{3\pi}{4}$ (departure from uniformity: $P < 0.01$, Rao's test).

23shown in Fig. 6B. While the phase in unseen trials was distributed uniformly around the circle (blue lines), in seen trials it concentrated around $\frac{\pi}{4}$ and $-\frac{3\pi}{4}$ in the two clusters, respectively. Phase $\frac{\pi}{4}$ and $-\frac{3\pi}{4}$ are opposite phases to each other, suggesting that the underlying dipole current flow is likely located between these two clusters of sensors in a perpendicular direction.

We next extracted two dominant clusters of sensors based on the correlation between phase (0.05 ~ 1Hz at stimulus onset) and subjective awareness for each subject in a similar manner (number of sensors included in each cluster: 12.8±2.8), and calculated their phase histograms for seen and unseen trials separately. From the phase histogram we obtained two metrics: i) a phase-locking value (PLV) describing the degree of phase concentration across trials, with value 0 indicating a uniform distribution and value 1 indicating identical phase value across trials; and ii) a preferred phase describing the most commonly observed phase. For the 22 clusters, PLV is plotted against the preferred phase for seen (red) and unseen (blue) trials respectively (Fig. 6C). The degree of phase concentration was significantly higher in seen trials than unseen trials (paired t-test on PLV across clusters, $P < 1.6e-12$). Moreover, in unseen trials, the distribution of preferred phase across the 22 clusters did not depart from uniformity (Rao's test for circular uniformity, $P > 0.5$), whereas the preferred phase in seen trials followed a bimodal distribution, concentrating around $\frac{\pi}{4}$ and $-\frac{3\pi}{4}$ (departure from uniformity: $P < 0.01$, Rao's test).

To ensure that the above results did not depend on our choice of clusters, we additionally performed a sensor-level analysis. All sensors from all subjects with a significant correlation between phase and subjective awareness (P < 0.001, assessed by ANOVA) were included in this analysis. We obtained the PLV and preferred phase for each sensor in seen and unseen trials respectively (Fig. 6D). Across 635 sensors, the preferred phase in seen trials again clustered around $\frac{\pi}{4}$ and $-\frac{3\pi}{4}$. The degree of phase concentration was dramatically higher in seen trials than unseen trials (paired t-test on PLV across sensors, P < 2.6e-298).

**Power of MEG Activity**

To investigate the relationship between the power of MEG activity and subjective awareness or objective performance, a three-way ANOVA (factors: SUB, OBJ and CONF) was carried out on MEG activity power from each sensor at every time-frequency location. In addition to the ANOVA, we also performed a two-sample t-test on power between seen and unseen trials, and between correct and incorrect trials. Because our previous analyses revealed widespread activity associated with subjective awareness (Figs. 3 & 4), for group analysis we adopted a whole-brain instead of region-of-interest approach. For subjective awareness and objective performance, the t-score was summed across sensors showing a significant ANOVA result (at a P < 0.05 level for SUB and OBJ respectively, for details see Materials and Methods). We then assessed whether this summed t-score ($\sum_T$) was significant across the 11 subjects (one-sample t-test against 0, corrected for multiple comparisons using cluster-based nonparametric permutation test). The results of this analysis are shown in Fig. 7. For the effect of subjective awareness (Fig. 7, left), we found that MEG power in α, β and low-γ frequency ranges (approximately 10 ~ 50 Hz) was lower in seen as compared with unseen trials, in a window of approximately 500 ms ~ 1 sec (cluster-level $P < 0.001$). By contrast, MEG power in the lowest frequencies (2 ~ 6 Hz) was significantly higher in seen trials than unseen trials during a 500-ms-long window following stimulus onset ($P < 0.001$). No significant result was observed in the high-γ frequency range. The correlation between MEG power and objective performance in the low-mid frequency ranges (<30 Hz) was qualitatively similar but much weaker (Fig. 7, right), with decreased power in α and β frequency ranges ($P < 0.008$) and increased power in the 2-Hz band ($P < 0.02$) in correct as compared with incorrect trials. In addition, correct trials were associated with increased power in the high-γ frequency range ($P < 0.008$).

**MEG Activity Correlated with Confidence**

Thus far we have focused on extracting brain activities underlying subjective awareness and objective performance while controlling for confidence as a covariate. Lastly, we investigated brain activities contributing to confidence. While subjective awareness and objective performance are both binary variables, confidence measure in our paradigm is an ordinal variable with four levels. Because ANOVA cannot extract MEG activity correlated with confidence in a monotonic manner, we adopted an alternative approach. After source localization, trial-by-trial ordinal regression (De Martino et al., 2013) was carried out at each source location to identify MEG activity monotonically correlated with confidence, with the effects of subjective awareness and objective performance both controlled for. The results of this analysis are shown in Fig. 8. Source activity correlated with confidence first appeared in the dorsal parietal cortex at around 200 ms. It then moved anteriorly and had widespread frontoparietal distribution at ~500 ms. The activity was restricted to posterior brain regions at ~750 ms and dissipated thereafter. It reappeared around central and frontal cortices at 2 ~ 2.5 sec, possibly in anticipation of the upcoming responses to questions. These results show that compared to subjective awareness (Fig. 4), confidence is associated with relatively transient MEG activity.

**Test-retest Reliability and Control for SNR across Behavioral Conditions**

Additional MEG data were collected for two control analyses. Firstly, in order to assess test-retest reliability, we obtained three experimental sessions on different days in two additional subjects (Subj. #12 and #13). A three-way ANOVA (factors: SUB, OBJ and CONF) was carried out on data from each session in the sensor space. Similar qualitative results were obtained across days in both subjects: namely, that long-lasting MEG activity correlated with subjective awareness but not objective performance (Fig. 9A). Because the sensor locations in relation to the brain varied across sessions, the spatial patterns of the activity were variable.

Secondly, because different behavioral conditions are associated with different numbers of trials (e.g., the three conditions in Fig. 2B contained 58, 32 and 24 trials, respectively), the differential SNR across conditions might confound our results. To address this issue, in Subj. #12 and #13, we conducted a control analysis by using an identical number of trials in each condition to compute the ERF. Results from a representative session in each subject are shown in Fig. 9B. These results are similar to those shown in Fig. 2B:

long-lasting MEG activity was observed only in "Seen and correct" condition, but not in "Unseen and correct" or "Unseen and incorrect" conditions. Thus, after controlling for SNR, we confirmed that slow MEG activity correlates with subjective awareness but not objective performance.

**Slow EEG Activity Correlates with Subjective Awareness**

Previous EEG studies have established that the negative shift of scalp-recorded SCPs indexes increased cortical excitability (Rockstroh et al., 1989; Birbaumer et al., 1990). Because the polarity of MEG signals depends on the position and direction of the underlying dipole current flow in relation to the MEG sensor, there is no unique relationship between MEG signal polarity and changes in cortical excitability. This is consistent with our finding that both positive- and negative- going slow MEG activities correlated with subjective awareness (Figs. 2 & 3). To further elucidate the underlying cortical excitability state associated with subjective awareness, we carried out a preliminary EEG study. The same two subjects as used in the additional MEG data collection (Subj. #12 and #13) each performed an EEG session under the same task paradigm. We sorted all trials into three categories as in Figs. 2B and 9B – "Seen and correct", "Unseen and correct", "Unseen and incorrect", and computed ERPs for each condition. The results are shown in Fig. 10. In the "Seen and correct" condition, we observed a transient negative potential at ~300 ms followed by long-lasting *positive* EEG potential that peaked around 500 ms and returned to baseline around 1.5 sec. The transient negative potential was distributed posteriorly, corresponding to the initial feedforward excitation in the early and higher-order visual cortices. The late positive potential had a central-parietal distribution, consistent with earlier studies (Dehaene and Changeux, 2011). This slow positive EEG potential suggests that it was profound cortical inhibition, instead of excitation, that correlated with subjective awareness of the stimulus.

**Behavioral Control for Working Memory Decay**

So far our paradigm has used a relatively long post-stimulus blank period (3 ~ 6 sec) in order to elucidate slow brain activity uncontaminated by behavioral responses. While this design is similar to several previous studies (e.g., Hesselmann et al., 2011), it introduces a potential behavioral confound of working memory decay during this period. To address this issue, we collected additional behavioral data in a cohort of 11 subjects using a 200-ms-long post-stimulus blank period under otherwise identical task design. We compared the behavioral results using this new data ("Short duration") with the original data under MEG recording (N = 11, "Long duration", same as reported in Fig. 1), both analyzed according to signal detection theory (Green and Swets, 1966) (Fig. 11).

For subjective awareness, the detection d' was highly significant for both long-duration and short-duration tasks ($P = 5e-7$ and $0.0004$, Fig. 11A). Moreover, subjects adopted a conservative criterion in both tasks, as reflected in the criterion being significantly above 0 ($P = 8e-7$ and $0.0008$, Fig. 11A). For objective performance, discrimination d' was also highly significant for both long-duration and short-duration tasks ($P = 8e-6$ and $5e-5$), while criterion was close to 0, indicating very small biases (Fig. 11B). A two-way repeated-measures ANOVA on d' [between-subject factor: task (long vs. short duration);

within-subject factor: behavior (subjective awareness/detection vs. objective performance/discrimination)] revealed a significant effect for task (P < 0.01) with higher d' in the long-duration task, while the effect of behavior and the interaction of duration × behavior were not significant. The higher d' during the long-duration task could be due to the smaller number of catch trials included therein, or better conscious/post-conscious processing allowed by the longer post-stimulus blank duration. A similar two-way ANOVA on criterion revealed a non-significant effect of task.

We next considered the discrimination d' for objective performance against subjective awareness and confidence and found a similar pattern across both tasks (Fig. 11C). The effect of subjective awareness was highly significant (Long duration: P = 4e-15; Short duration: P = 3e-9), and the interaction of subjective awareness and confidence was significant (Long duration: P = 0.01; Short duration: P = 0.0006), while confidence by itself was not significant in either task (P > 0.7). Altogether, these results suggest that the behavioral results we obtained under MEG recording were qualitatively similar to those obtained in the behavioral control experiment with a much shorter post-stimulus blank period (200 ms vs. 3~6 sec) and therefore were unlikely to be significantly affected by potential working memory decay during the post-stimulus period.

**DISCUSSION**

In summary, we found that long-lasting, low-frequency MEG activity correlates with the state of subjective awareness, even after the effects of objective performance and confidence were both controlled for. In the time domain, this activity lasts for up to 2 sec after the onset of a brief stimulus (<67 ms long). In the frequency domain, the phase and power of the lowest frequencies (<5 Hz) correlate with subjective awareness. By contrast, objective performance and confidence are both contributed by relatively transient brain activity. These results provide strong support for the SCP hypothesis on conscious processing (He and Raichle, 2009b). Lastly, source modeling suggests that widespread frontoparietal and temporal cortical areas contribute to the SCPs underlying subjective awareness.

**Connection to Previous Literature on Conscious Perception**

Our results are consistent with the global neuronal workspace (GNW) theory of consciousness in terms of emphasizing late (>200 ms) activity and contributions from widespread brain areas (Dehaene and Changeux, 2011). Our findings differ from the GNW framework in several aspects. First, experimental work supporting the GNW theory has emphasized brain activity in the time window of 200 ~ 600 ms (Sergent et al., 2005; Del Cul et al., 2007), while the long-lasting activity we observed persisted for ~2 sec after stimulus offset. This is potentially due to the fact that in these prior studies, a high-pass filter at around 0.5 Hz was applied (as opposed to 0.05 Hz used in the current study), such that the very slow activity was lost. Secondly, studies under the GNW framework have generally interpreted the late EEG activity as part of the P3b potential (Sergent et al., 2005; Del Cul et al., 2007; Lamy et al., 2009). Experimentally, this is consistent with the slow positive potential in seen trials observed in our EEG recordings (Fig. 10). However, our interpretation of these results departs from the GNW theory. We think that the slow positive potential and the related P3b potential are part of the SCP family, indicating a reduction in cortical excitability (Birbaumer and Elbert, 1988; Deecke and Lang, 1988; He and Raichle, 2009a). Thus, instead of "global ignition" or "broadcasting", we believe that these data reveal global inhibition related to the updating of conscious content/working memory and simultaneous inhibition of irrelevant information when a particular stimulus reaches conscious awareness (Birbaumer and Elbert, 1988).

Our results showing that SCPs correlate with conscious perception are consistent with an earlier study reporting that elevated neuronal firing in the medial temporal lobe can persist for ~2 sec after the recognition of a brief stimulus in a visual backward masking paradigm (Quiroga et al., 2008). Such long-lasting brain activity underlying conscious perception provides a potential mechanism for the well-documented effect that conscious experience can be modulated up to hundreds of milliseconds after stimulus offset or motor output by external events, as in flash-lag illusion (Eagleman and Sejnowski, 2000), post-stimulus cortical stimulation (Libet, 1982) or attentional cueing (Sergent et al., 2013), and post-movement transcranial magnetic stimulation (Lau et al., 2007).

A previous study reported that the phase of spontaneous SCP recorded by EEG over occipital cortex modulates conscious visual threshold, such that liminal stimuli presented on the negative SCP shift were more likely to be perceived (Devrim et al., 1999). This result is consistent with a large literature showing that the negative shift of SCPs indexes increased cortical excitability (Rockstroh et al., 1989; Birbaumer et al., 1990). At present the phase relationship between SCPs recorded by MEG and those recorded by EEG remains unknown. We observed that across many sensors phase $\frac{\pi}{4}$ and $-\frac{3\pi}{4}$ of slow (<1 Hz) MEG activity at stimulus onset predicted a higher probability of subjectively perceiving the stimulus (Fig. 6). Phase $\frac{\pi}{4}$ and $-\frac{3\pi}{4}$ are opposite phases to each other, consistent with the fact that MEG sensors located on opposite sides of a dipole record activities that are mirror images of each other with opposite signs. Future simultaneous EEG-MEG studies will be needed to elucidate the phase relationship between EEG- and MEG- recorded SCPs.

A recent study using a whisker stimulus detection task in rodents found more depolarized membrane potentials in primary sensory neurons at 100~300 ms following stimulus onset during hit than miss trials (Sachidhanandam et al., 2013). This result corresponds well with our EEG finding of a negative potential at ~300 ms over posterior visual regions (Fig. 10). Nonetheless, whole-brain EEG allowed us to observe the ensuing widespread positive potential over frontoparietal cortices. In addition, subjective report in human subjects avoids the ambiguity of using detection performance to infer the state of subjective perception.

Increased power in the gamma-frequency range was previously reported to correlate with conscious perception (Fisch et al., 2009; Gaillard et al., 2009; Wyart and Tallon-Baudry, 2009). By contrast, we found gamma-frequency power to correlate with objective performance but not subjective awareness (Fig. 7). A major distinction between the current work and these previous studies is that we controlled for objective performance and confidence. Consistent with our finding, a recent study found localized gamma-frequency power to be correlated with sensory evidence but not conscious perception (Aru et al., 2012b).

**Connection to Previous Literature on Metacognitive Confidence**

We found that after the effects of subjective awareness and objective performance were controlled for, early (~200 ms) MEG activity in the dorsal parietal cortex and late (~2.5 sec) MEG activity in the anterior prefrontal cortex both correlated with subjects' confidence level (Fig. 8). These results are respectively consistent with prior primate neurophysiology (Kiani and Shadlen, 2009) and human functional/anatomical MRI (De Martino et al., 2013; McCurdy et al., 2013) studies. Our findings suggest that the different brain regions uncovered by neurophysiology and fMRI might simply reflect the difference in temporal sensitivities of these methods.

**Mechanisms of SCPs**

Early neurophysiological studies suggested: "Simultaneous recording of membrane potentials, extracellular and intracellular recording from apical dendrites, and field potentials clearly demonstrate long-lasting EPSPs at the apical dendrites as the main factor underlying negative SCPs" (Birbaumer et al., 1990). However, to date the source of these long-lasting EPSPs remains unclear. Nonetheless, there are several plausible and not mutually exclusive candidates. First, recurrent excitatory neuronal networks may constitute a prominent source of long-lasting EPSPs (Major and Tank, 2004; Wong and Wang, 2006; Chaudhuri et al., 2014). Second, long-lasting cellular mechanisms such as calcium spikes (Larkum, 2013), metabotropic receptors (Zhang and Seguela, 2010) and the endocannabanoid pathway (Carter and Wang, 2007) could potentially produce long-lasting EPSPs in the SCP time-scale.  Third, it has been long conjectured that neural modulations such as acetylcholine might play a role in generating SCPs (Birbaumer et al., 1990).  A recent study demonstrating that cholinergic inputs to layer I could selectively disinhibit layer II/III pyramidal neurons lends substantial credence to this idea (Letzkus et al., 2011).  Future experimental investigations should test and refine these predictions.  A promising avenue is to combine DC-recordings of field potentials (Kahn et al., 2013; Pan et al., 2013) with pharmacological manipulations or stimulation of specific cell types in different cortical layers.

**Caveats and Future Questions**

Our results suggest that the waveform, phase and power of slow MEG activity correlate with the state of subjective awareness from trial to trial, after objective performance and confidence were both controlled for.  These results provide strong support for the SCP hypothesis on conscious processing (He and Raichle, 2009b).  Nonetheless, it is important to clarify that not all SCPs are directly related to conscious processing.  For example, unconscious modulation of brain excitability can also manifest as SCPs (Libet et al., 1983; Elbert, 1990).  This is not in conflict with the SCP hypothesis on conscious processing – as we stated in the original paper (He and Raichle, 2009b), "These results, however, do not suggest that the negative SCP, whenever it appears, is an index of conscious awareness."  A crucial goal for future research is to elucidate the difference in spatiotemporal patterns of the SCPs related to unconscious vs. conscious processing.

Secondly, the relationship between our results and various forms of memory should be explored in future experiments. It is unlikely that visual persistence or iconic memory confounds our results. Visual persistence lasts up to several hundred milliseconds and iconic memory up to a second – both are much shorter than our SCP findings (Sperling, 1960; Coltheart, 1980). Moreover, both visual persistence and iconic memory are strongest right after stimulus offset and decay in a monotonic fashion; by contrast, we found that SCPs correlated with subjective awareness reach their peak at ~500 ms and decay thereafter (Figs. 2C, 3 & 10). Nonetheless, the potential roles of fragile visual short-term memory (FM) and working memory (WM) should be further investigated in future studies. Existing evidence suggests that FM is a form of location- and object-specific short-term memory store that is likely mediated by higher-order visual areas such as V4 and inferotemporal cortex (Sligte et al., 2009; Pinto et al., 2013). As such, neural processes underlying FM might manifest in the negative SCP potential peaking at 300 ~ 400 ms localized to posterior visual regions that was observed in our EEG experiment (Fig. 10). Moreover, as discussed earlier, the widespread cortical inhibition thereafter, indicated by slow positive potentials at >500 ms (Fig. 10), might be related to the updating of WM when a stimulus reaches conscious awareness. At present, it remains unclear whether these – conscious access, FM, and WM update – are separable processes (e.g., see discussions in Sligte et al., 2008). In addition, it is possible that delaying responses in our paradigm introduced a sustained WM component (e.g., Vogel and Machizawa, 2004); future studies dissociating these processes will be valuable.

Lastly, the separation of the NCC from NCC-pr and NCC-co in our study is likely incomplete. Our analyses were able to disentangle these phenomena to the extent that objective performance captures NCC-pr and metacognitive confidence captures NCC-co. However, there are likely NCC-pr and NCC-co processes not captured by these measures. For example, it has long been postulated that conscious perception could lead to verbal report or the formation of long-term memory (Baars, 1988). Although our paradigm did not involve overt verbal report, and the effect of long-term memory was likely mitigated by repeated presentations of a minimal stimulus, at present it is not possible to rule out latent processes as such. These caveats notwithstanding, the current study constitutes a significant step forward in the ongoing effort to disentangle conscious from unconscious brain processes and distill the NCC from NCC-pr and NCC-co.

## ACKNOWLEDGEMENTS

This research was supported by the Intramural Research Program of the NIH/NINDS. We thank Tom Holroyd, Ziad Saad, Gang Chen, John Ostuni and Ole Jensen for helpful discussions.

**Figure Legends**

**Figure 1. Task paradigm and behavioral results.** (**A**) In each trial, a Gabor patch with two possible orientations (45° or 135°) is shown for a brief duration, sandwiched between two blank screens. The duration distribution of the second blank screen from an example subject is shown in the inset. At the end of each trial, the subject is prompted to answer three sequential questions at their own pace. (**B**) The total fractions of trials in which the subjects reported seeing the stimulus ("seen" answers to the $2^{nd}$ question) or correctly discriminated the orientation of the Gabor patch (correct answers to the $1^{st}$ question). (**C**) The fraction of correct trials (based on orientation discrimination) in "seen" and "unseen" trials respectively. (**D**) Objective performance (based on orientation discrimination) conditioned on subjective awareness ("seen" vs. "unseen") and confidence level. (**E**) Subjective awareness conditioned on objective performance (correct vs. incorrect) and confidence level. In **B-E**, results are averaged across subjects, with error bars denoting s.e.m. across subjects.

**Figure 2. ERFs.** (**A**) Data from Subj. #1. Top: All trials were sorted into "seen" and "unseen" groups, and averaged separately. Bottom: All trials were sorted by whether the orientation of the Gabor patch was correctly identified; correct and incorrect trials were averaged separately. Different traces indicate different MEG sensors. Four sensors are highlighted in thicker lines for visualization purpose. (**B**) Data from Subj. #1. All trials were sorted into four categories: seen and correct (top), unseen and correct (middle), unseen and incorrect (bottom), seen and incorrect (not shown, due to the small number of trials), and averaged separately. The same four sensors as in **A** are highlighted. (**C**) Grand average on the absolute value of the MEG activity time course across all sensors and all subjects, for the three task conditions separately: Seen and correct (red); Unseen

and correct (blue); Unseen and incorrect (green). Shaded areas indicate s.e.m. across subjects.

**Figure 3. Trial-by-trial three-way ANOVA (factors: OBJ, SUB and CONF) in the sensor space.** (**A**) Results from the ANOVA for subjective awareness (left column) and objective performance (right column), respectively. Three example subjects are shown. For each subject, the ANOVA result from each MEG sensor is shown in a separate row. Color indicates P-values for the effect of SUB or OBJ, with warm colors indicating higher MEG activity in seen or correct trials, and cool colors indicating higher MEG activity in unseen or incorrect trials. (**B**) The number of significant sensors correlated with subjective awareness (blue) or objective performance (red) ($P < 0.05$, three-way ANOVA) averaged across all 11 subjects. Shaded areas denote s.e.m. across subjects. Red dots indicate the time points at which the two curves significantly depart ($P < 0.05$, paired t-test across subjects). (**C**) The P-values of the paired t-test comparing the two curves in **B**. Red dashed line indicates the $P = 0.01$ level. (**D-E**) As in B, except that significant sensors were determined by a threshold of $P < 0.01$ (**D**) or $P < 0.001$ (**E**).

**Figure 4. Trial-by-trial three-way ANOVA (factors: OBJ, SUB and CONF) in the source space.** At each source location, color indicates the percentage of subjects showing a significant ($P < 0.05$) effect of subjective awareness (Sub) or objective performance (Obj). Only source locations with at least 4 out of 10 subjects showing a significant result are included (corresponding to population-level $P < 0.001$, uncorrected). Ten time points from 100 ms to 2.5 sec after stimulus onset are shown.

**Figure 5. Trial-by-trial two-way circular ANOVA (factors: OBJ, SUB) on phase in the sensor space.** (**A**) Results from the ANOVA for subjective awareness (left column) and objective performance (right column) respectively. Results from 3 example subjects are shown. Color plots the number of significant (P < 0.001) sensors at each time-frequency location. (**B**) A paired t-test across all 11 subjects was performed at each time-frequency location on the number of sensors showing a significant (P < 0.001) SUB vs. OBJ effect. The P-values are plotted as color and thresholded at P < 0.01.

**Figure 6. Phase of low-frequency MEG activity in seen vs. unseen trials.** (**A**) Scalp plot of P-values for the effect of subjective awareness determined by the two-way ANOVA on phase (0.05 ~ 1 Hz activity at stimulus onset), from an example subject (Subj. #1). Two dominant sensor clusters are defined by the contour of P-values (P < 0.0001, indicated by *). (**B**) The phase histograms for seen (red) and unseen (blue) trials for the two dominant clusters in **A**. (**C**) Phase-locking value (PLV) is plotted against preferred phase for the two dominant clusters from each subject, calculated for seen (red) and unseen (blue) trials separately. The numbers indicate the cluster number (from 1 to 22). (**D**) PLV and preferred phase for all significant (P < 0.001, from ANOVA result for SUB effect) sensors in seen (red) and unseen (blue) trials respectively. The blue sinusoidal curves at the bottom of **C** and **D** show that phase 0 relates to the peak of the fluctuation, and phase ±π to the trough of the fluctuation.

**Figure 7. Group analysis results on MEG power for the effect of subjective**

**awareness (left) and objective performance (right).** For each subject and at each time-frequency location, the t-scores from a two-sample t-test (seen vs. unseen, or correct vs. incorrect) were summed across significant sensors as determined by the ANOVA ($\sum_T$, see Materials and Methods). The $\sum_T$ value was subjected to a population test across subjects (one-sample t-test against 0), and the resulting t-score is plotted as color in the top panels. Warm colors indicate that power was higher in seen (left) or correct (right) trials; cool colors indicate that power was higher in unseen or incorrect trials. Bottom: Significant time-frequency clusters after controlling for multiple comparisons using a nonparametric permutation test. The statistical significance of each cluster is indicated in the graph; color is the same as in top panels, indicating the population-test t-score.

**Figure 8. MEG activity correlated to confidence in the source space.** At each source location, the relationship between estimated source activity and confidence level was evaluated by ordinal regression (covariates: SUB and OBJ). Only source locations with at least two subjects showing a significant result were included. The agreement coefficient (see Materials and Methods) is plotted as color at different post-stimulus time points.

**Figure 9. Test-retest reliability assessment and control for SNR.** (**A**) Sensor-space three-way ANOVA (factors: SUB, OBJ and CONF) was carried out for each recording session in Subj. #12 and #13. The effect of SUB and OBJ are shown in the left and right columns, respectively. Format is the same as in Fig. 3A. (**B**) The ERFs from a representative session in each subject (day 2 in Subj. #12; day3 in Subj. #13) for three

conditions separately: "Seen and correct", "Unseen and correct" and "Unseen and incorrect". The number of trials was equated across the three conditions by dropping out a random fraction of trials, resulting in 84 trials per condition in Subj. #12 and 76 trials per condition in Subj. #13. Format is the same as in Fig. 2B.

**Figure 10. DC-EEG data from Subj. #12 (A) and Subj. #13 (B).** ERPs from all electrodes are shown for three conditions: "Seen and correct" (middle row), "Unseen and correct" and "Unseen and incorrect" (bottom row). The numbers of trials included in each condition are indicated in the figure. The scalp topography of voltage distribution at selected time points for the "Seen and correct" condition are shown in the top row.

**Figure 11. Behavioral control for the duration of post-stimulus blank period.** Behavioral data from the MEG sessions ("Long duration", same as data used in Fig. 1) and from behavioral testing in an additional cohort of 11 subjects using a 200-ms post-stimulus blank period ("Short duration") are analyzed according to signal-detection theory. (**A**) Detection d' and criterion c for subjective awareness. (**B**) Discrimination d' and criterion c for objective performance. (**C**) Discrimination d' under different states of subjective awareness (seen vs. unseen) and confidence level. In all graphs, error bars denote s.e.m. across subjects.

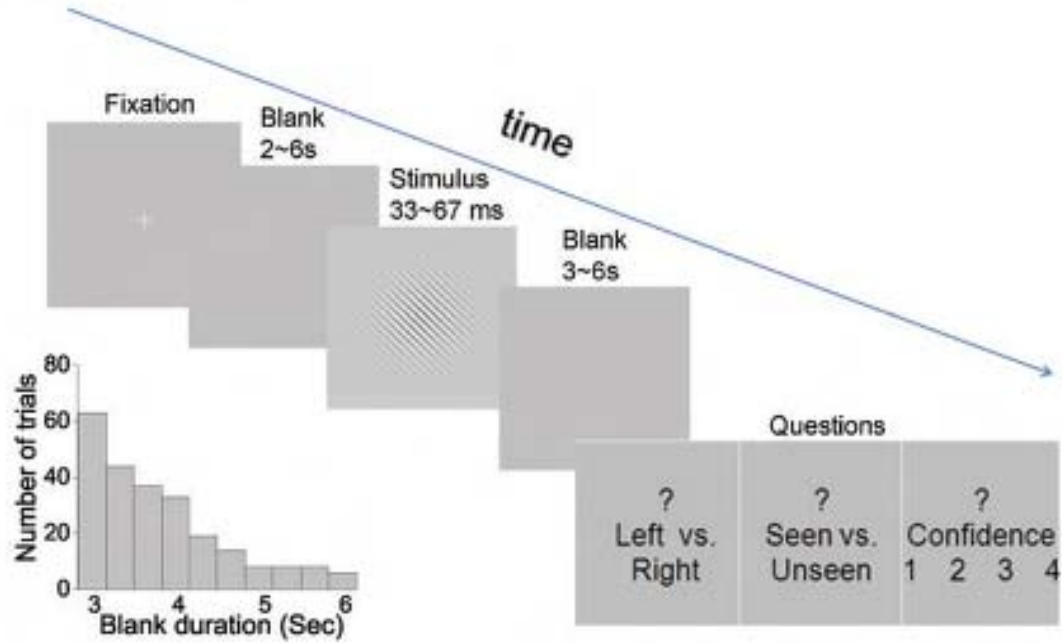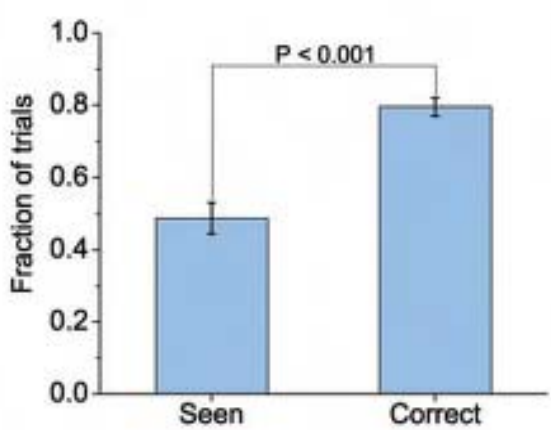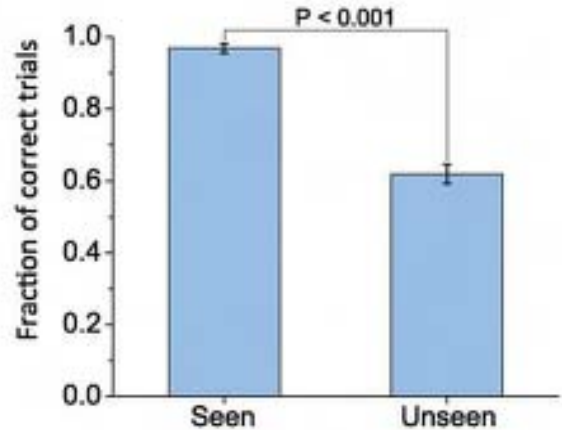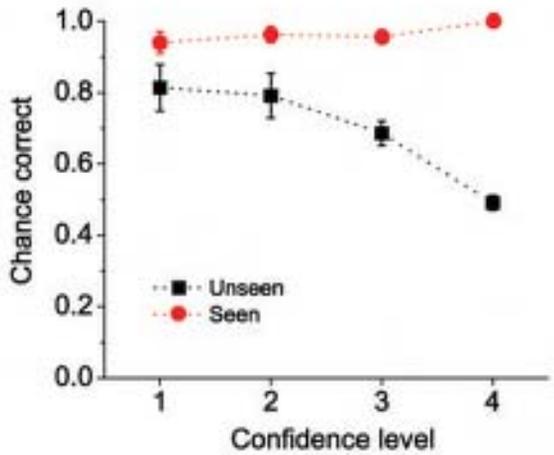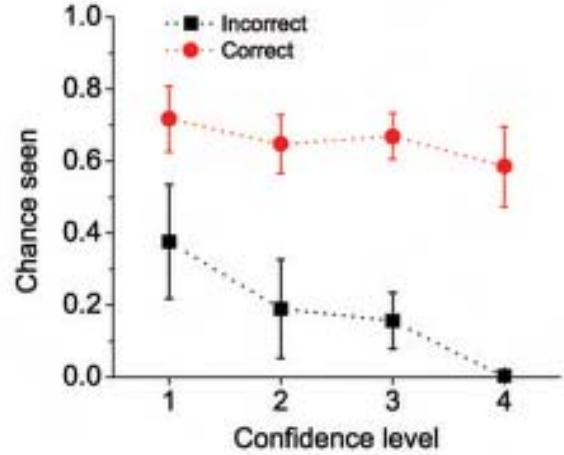

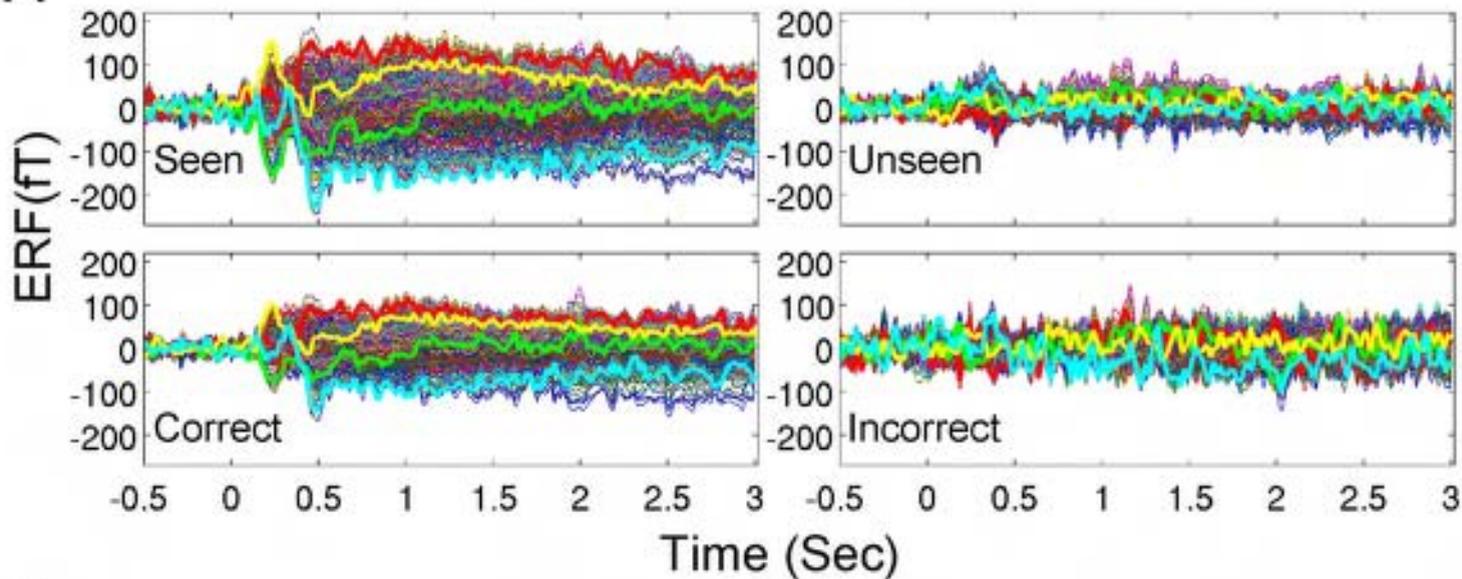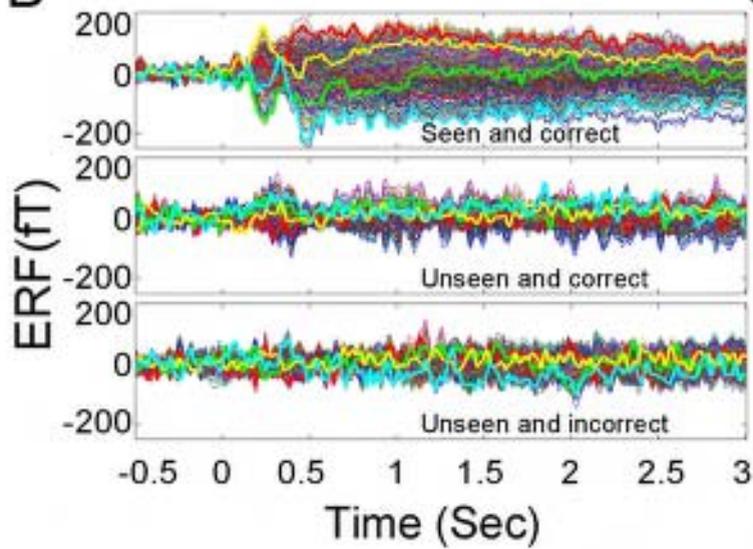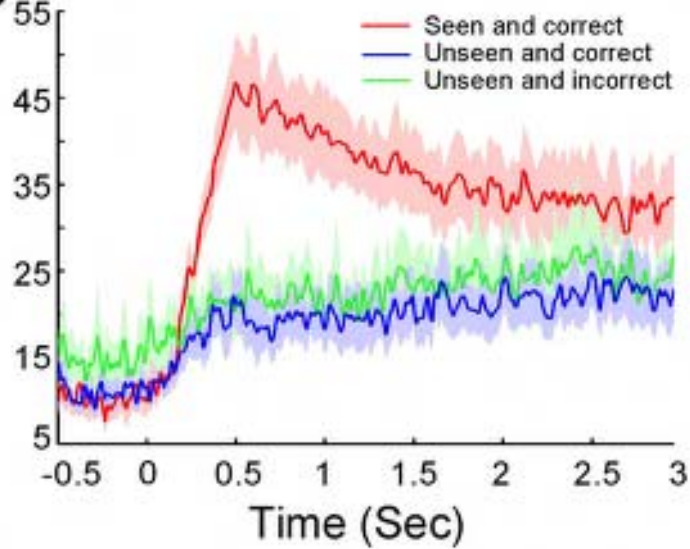

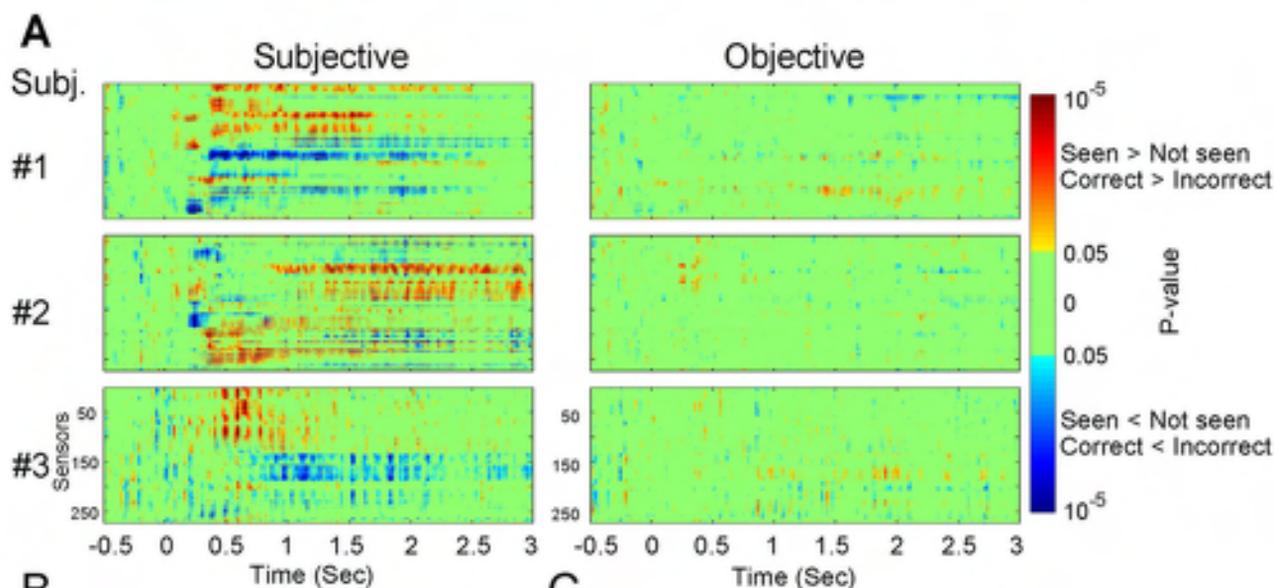

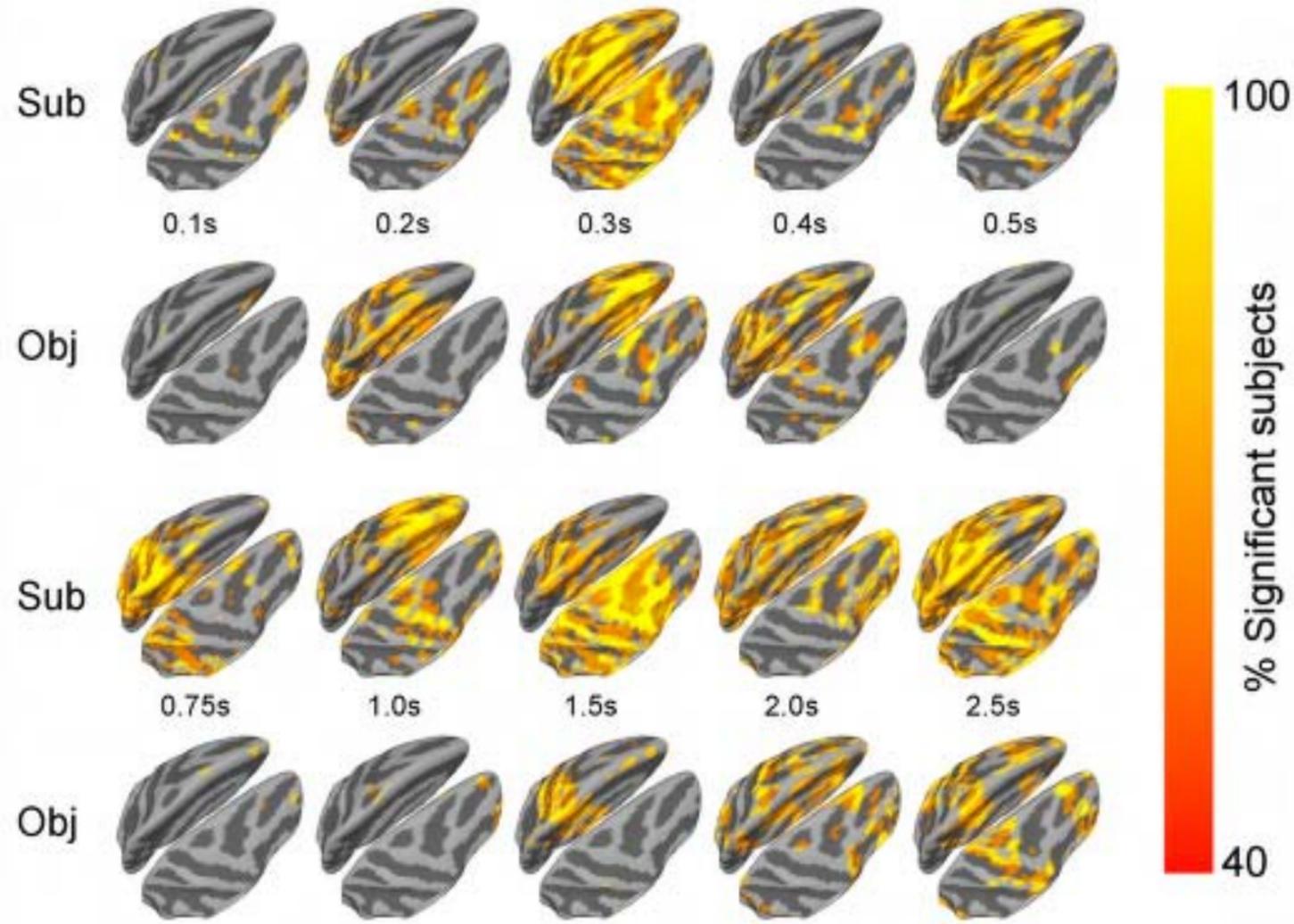

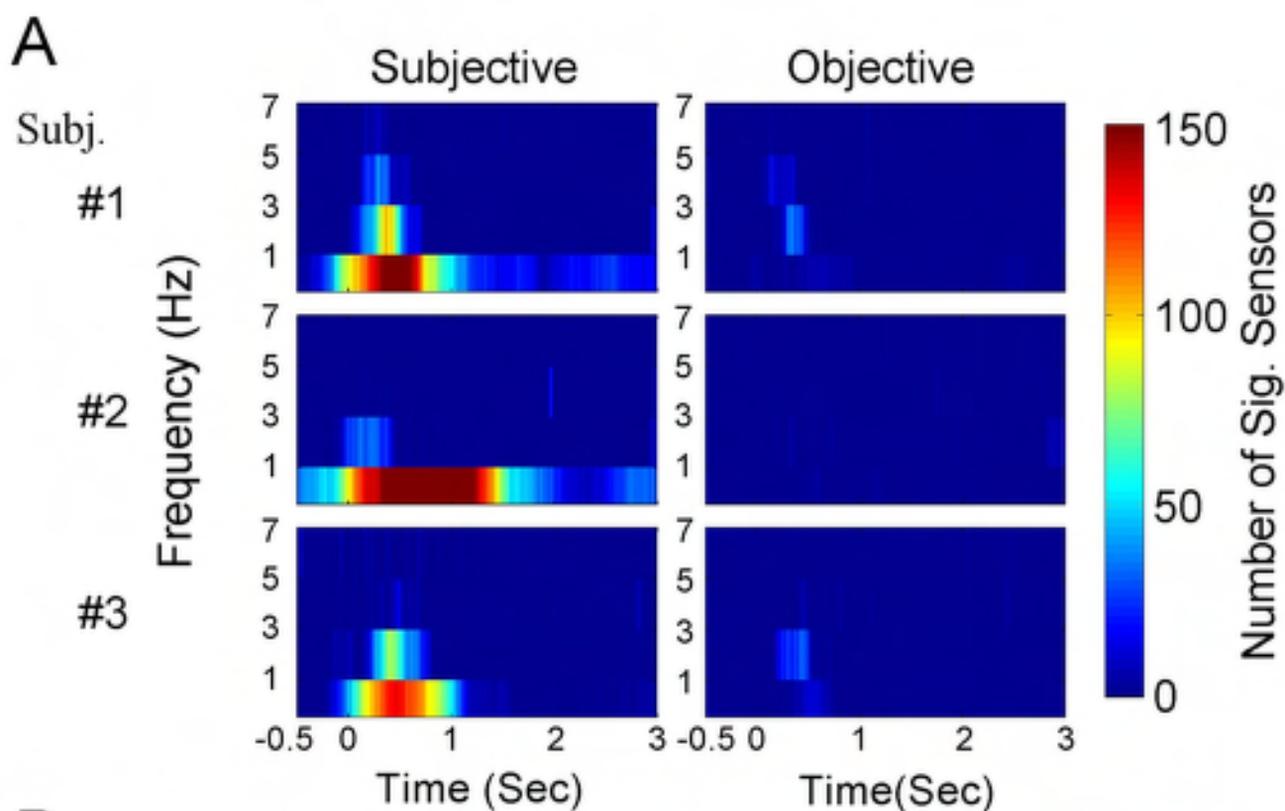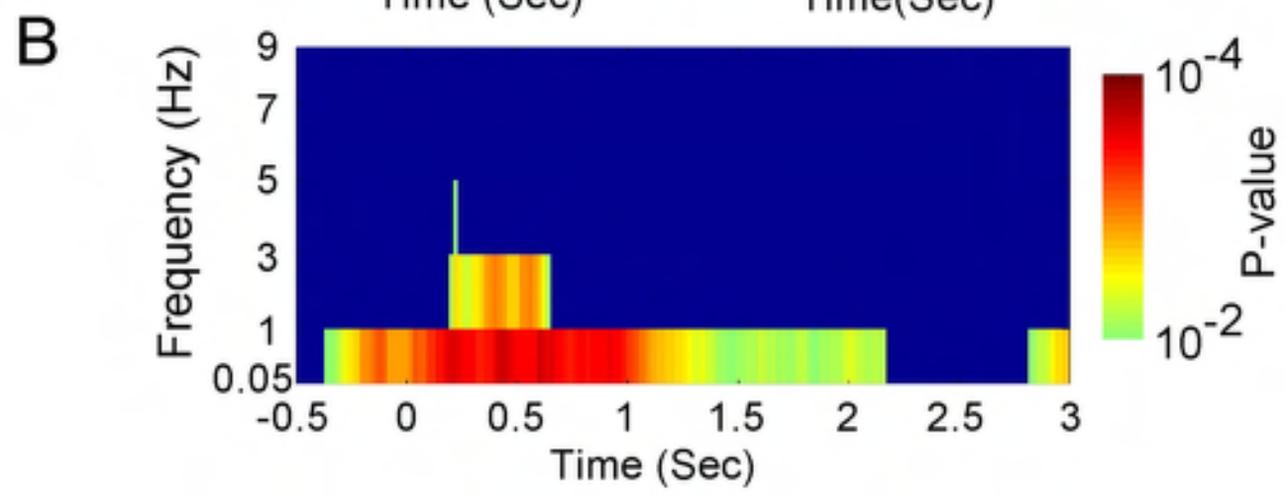

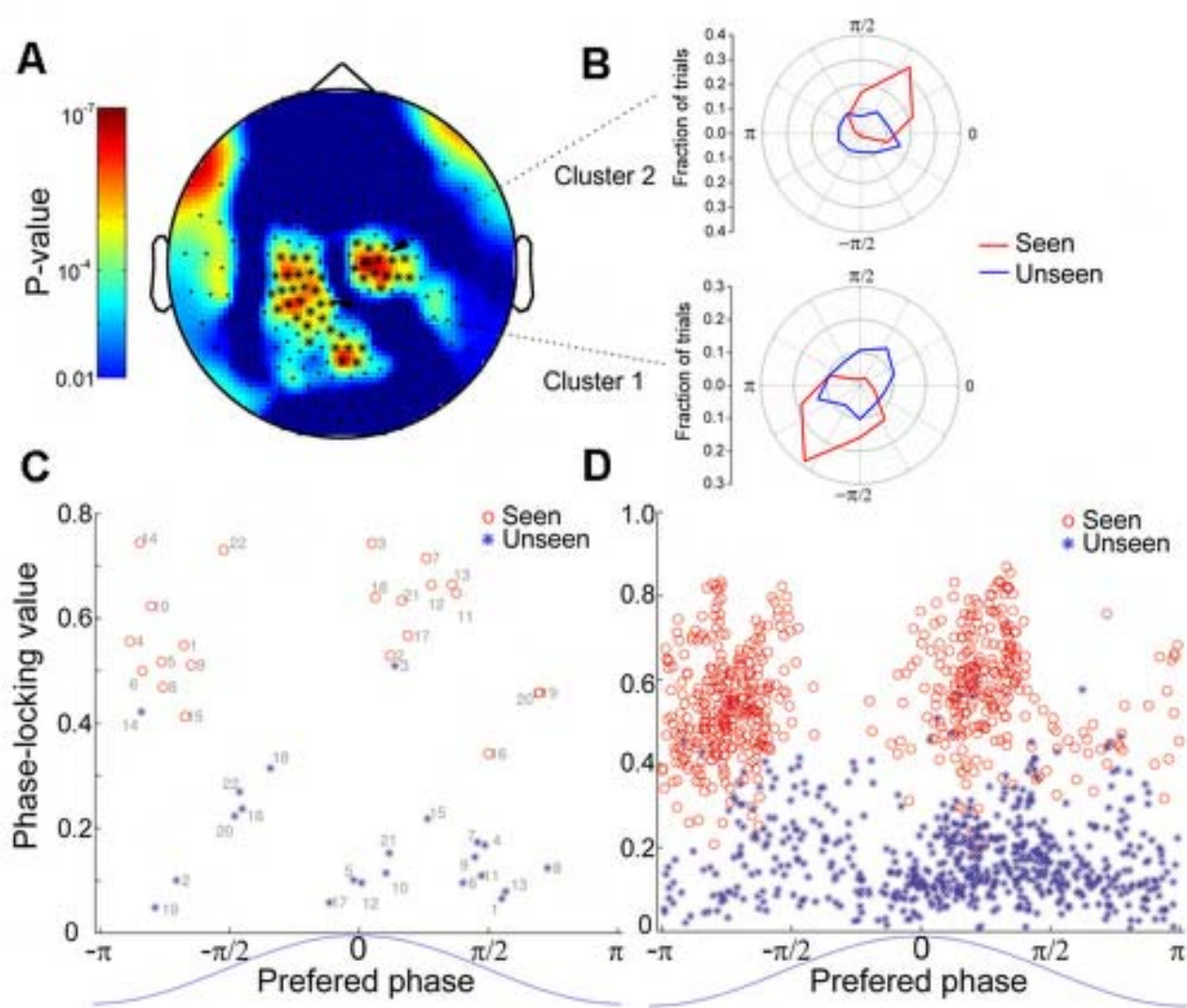

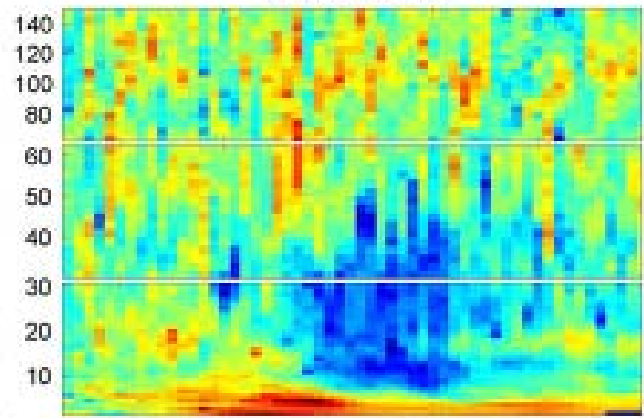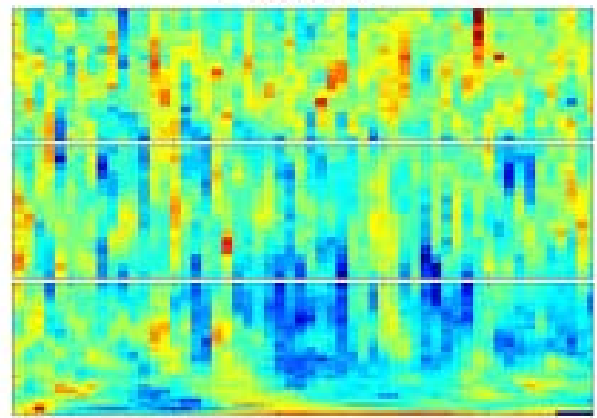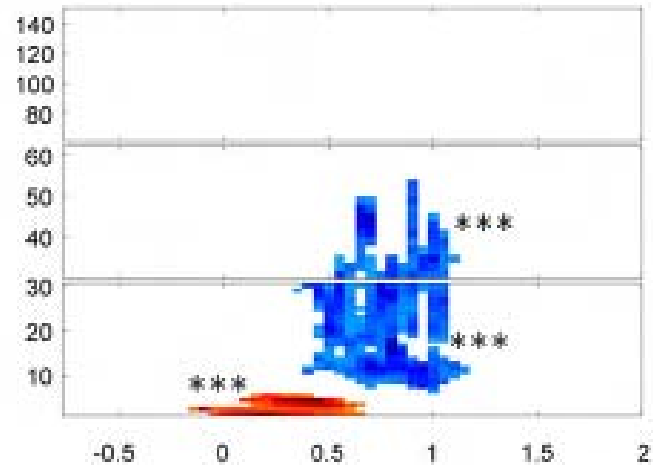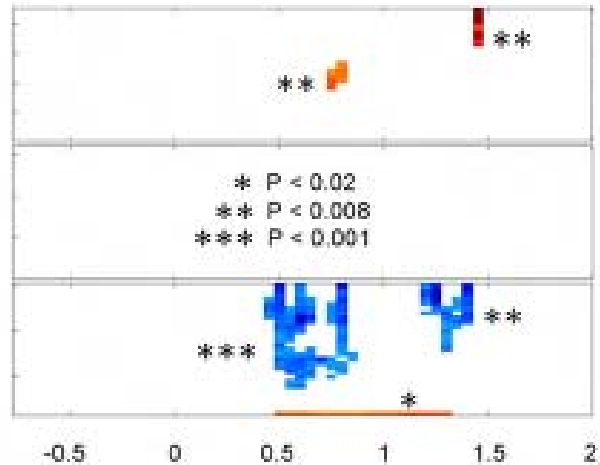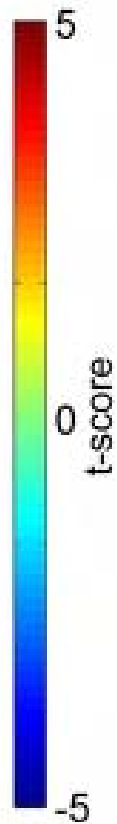

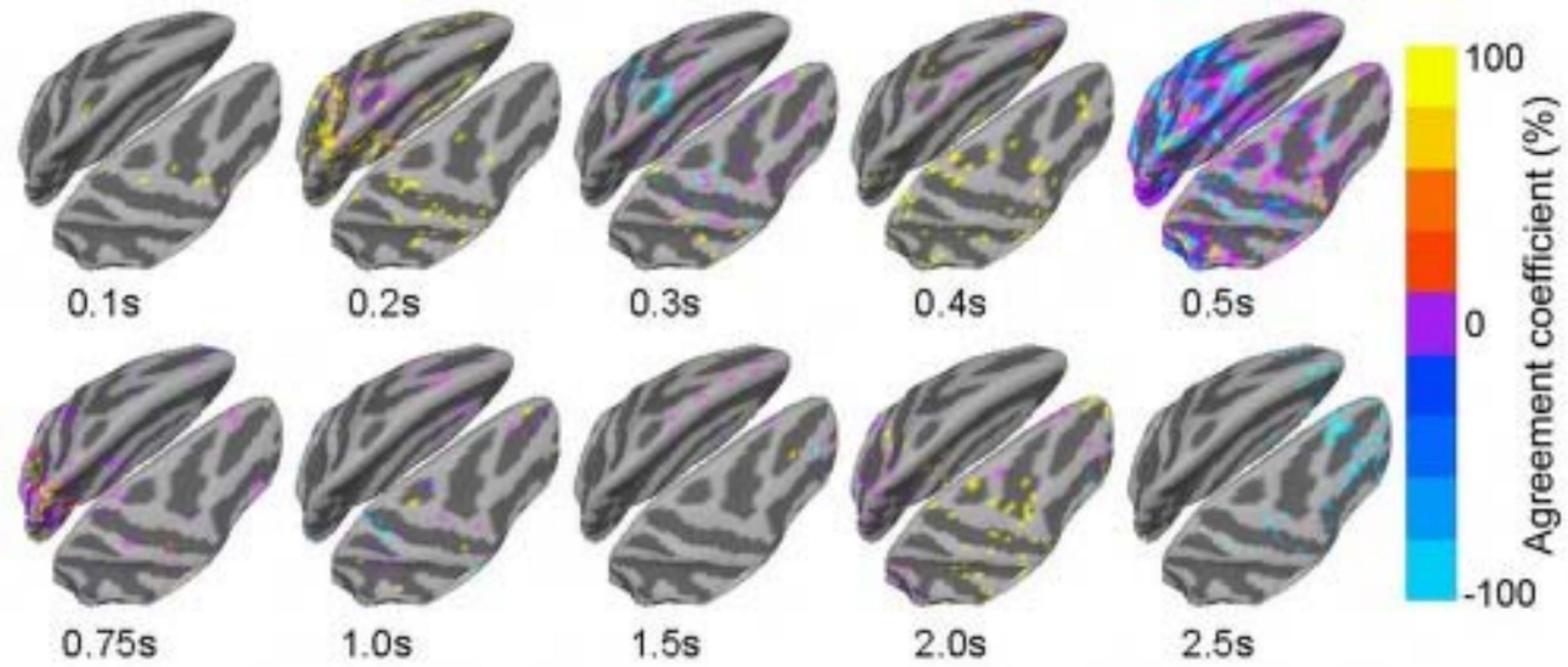

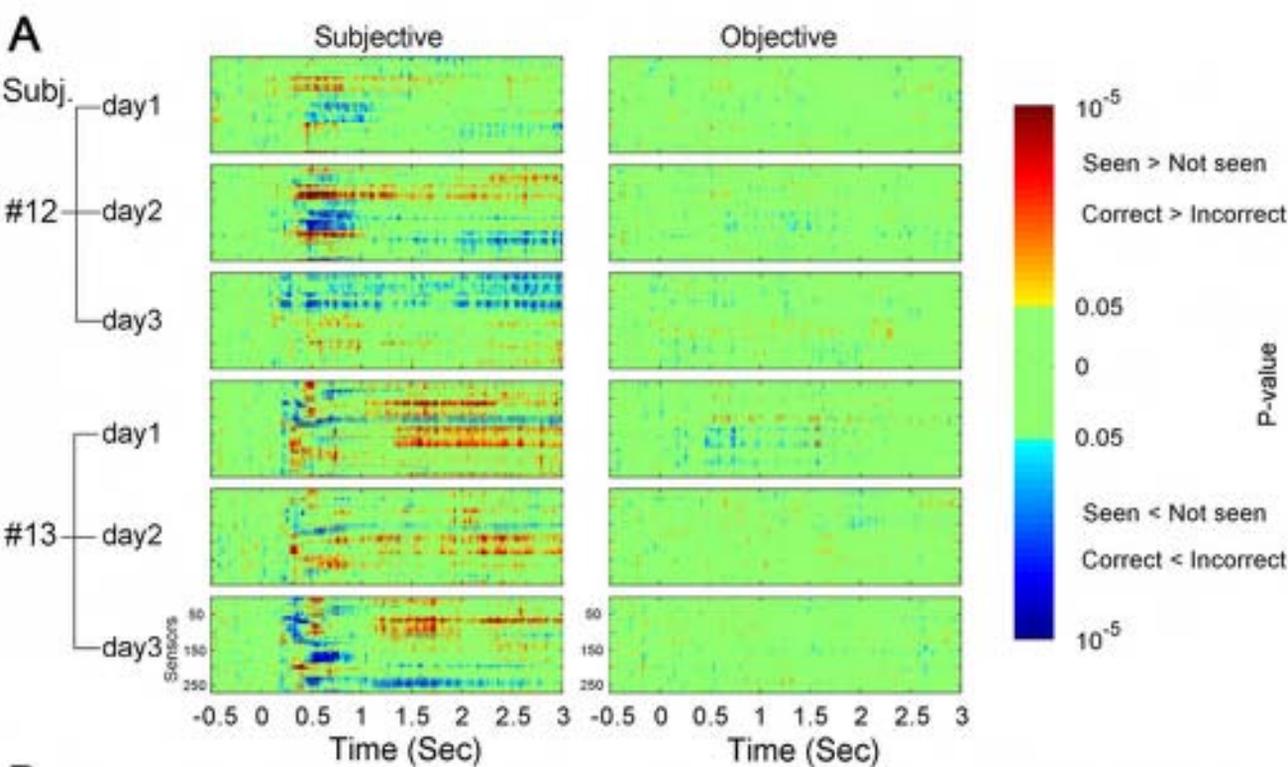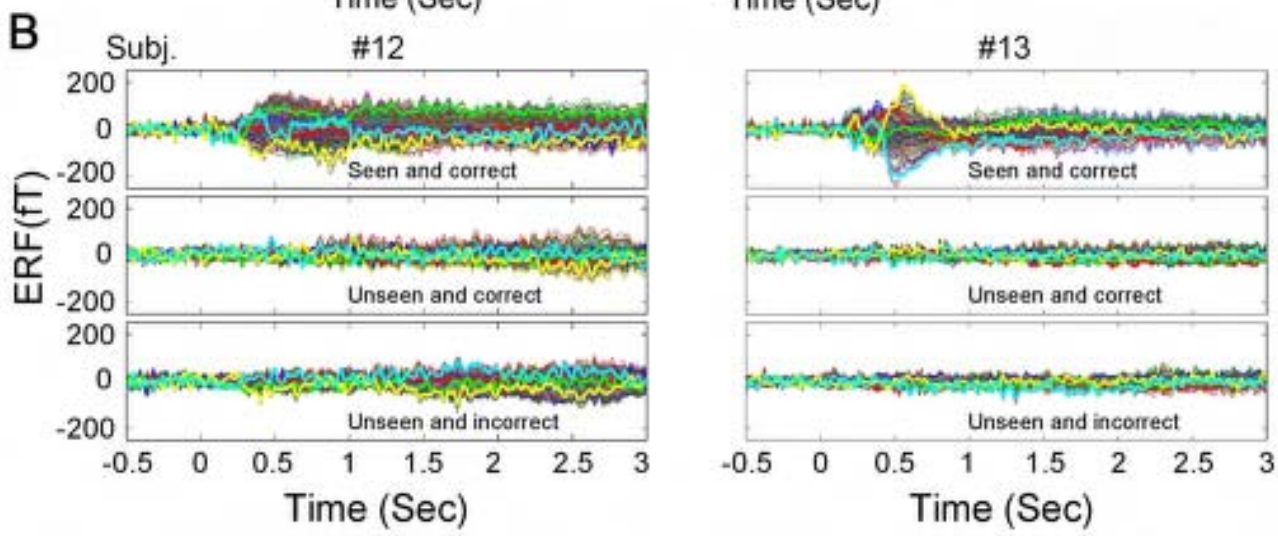

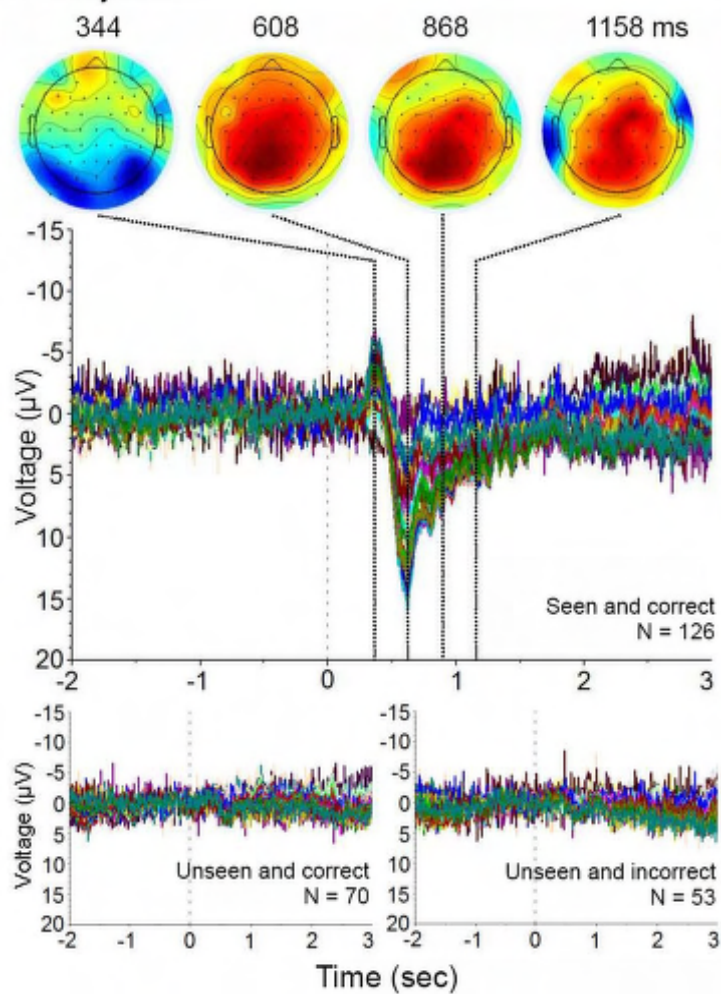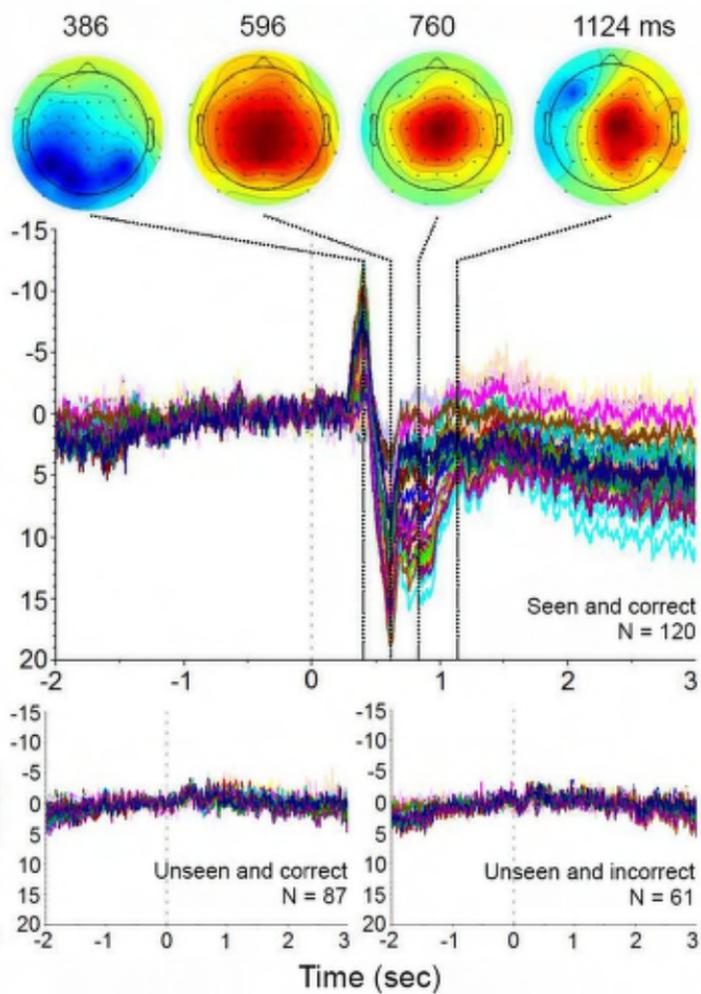

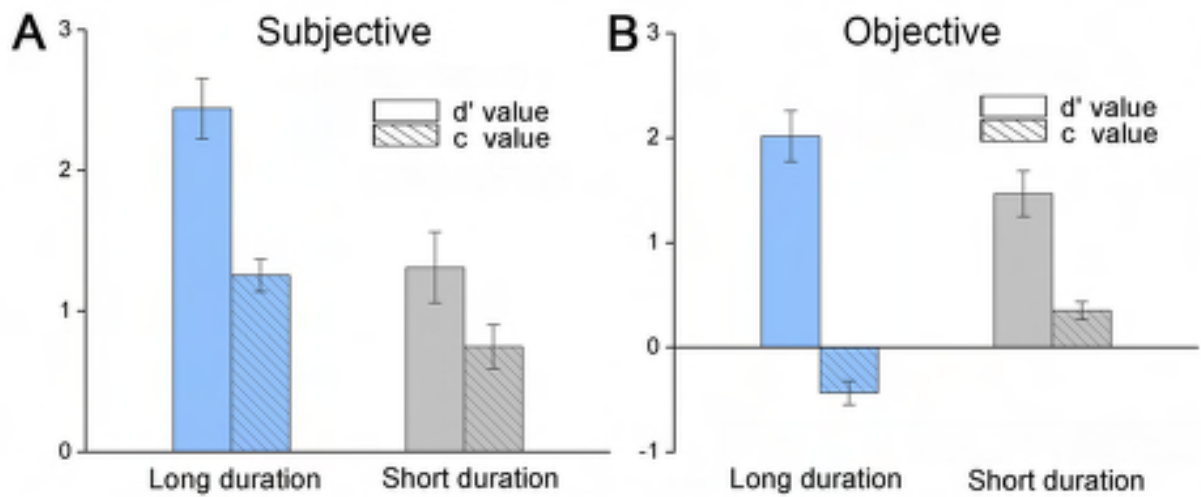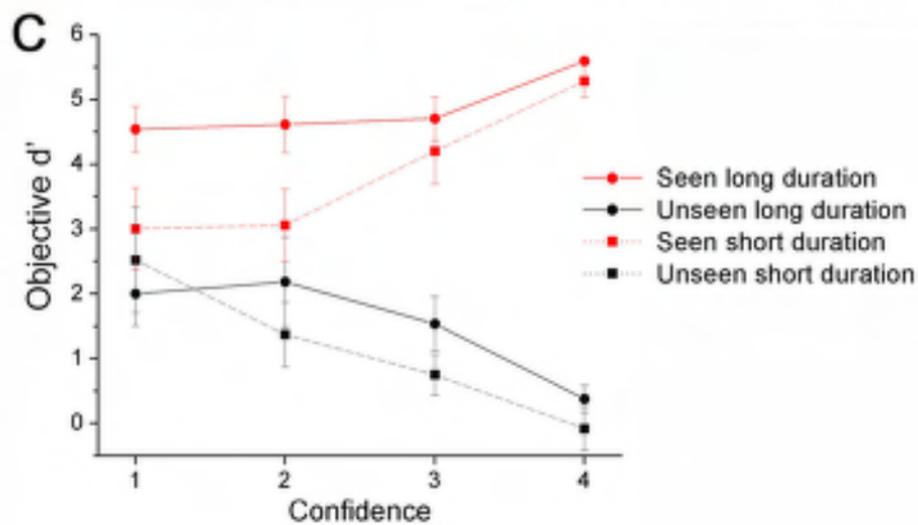